%% file: 0_main.tex
\documentclass[10pt,conference]{IEEEtran}
\IEEEoverridecommandlockouts
\usepackage{cite}
\usepackage{amsmath,amssymb,amsfonts}
\usepackage{algorithmic}
\usepackage{graphicx}
\usepackage{textcomp}
\usepackage{xcolor}
\usepackage{todonotes}
\usepackage{float}
\usepackage[hidelinks]{hyperref}
 \usepackage[T1]{fontenc}

\def\BibTeX{{\rm B\kern-.05em{\sc i\kern-.025em b}\kern-.08em
    T\kern-.1667em\lower.7ex\hbox{E}\kern-.125emX}}
\begin{document}

\title{Pinpointing Anomaly Events in Logs from Stability Testing  -- N-Grams vs. Deep-Learning

\thanks{We acknowledge the funding by Academy of Finland (grant ID 328058)}
}

\author{\IEEEauthorblockN{Mika Mäntylä}
\IEEEauthorblockA{\textit{M3S, ITEE, University of Oulu} \\
Oulu, Finland \\
mika.mantyla@oulu.fi}
\and
\IEEEauthorblockN{Martín Varela}
\IEEEauthorblockA{\textit{Profilence} \\
Oulu, Finland \\
martin.varela@profilence.com}
\and
\IEEEauthorblockN{Shayan Hashemi}
\IEEEauthorblockA{\textit{M3S, ITEE, University of Oulu} \\
Oulu, Finland \\
shayan.hashemi@oulu.fi}

}
\maketitle

\begin{abstract}

As stability testing execution logs can be very long, software engineers need help in locating anomalous events. 
We develop and evaluate two models for scoring individual log-events for anomalousness, namely an N-Gram model and a Deep Learning model with LSTM (Long short-term memory).  Both are trained on normal log sequences only. We evaluate the models with long log sequences of Android stability testing in our company case and with short log sequences from HDFS (Hadoop Distributed File  System) public dataset. We evaluate next event prediction accuracy and computational efficiency.
The LSTM model is more accurate in stability testing logs (0.848 vs 0.865), whereas in HDFS logs the N-Gram is slightly more accurate (0.904 vs 0.900). The N-Gram model has far superior computational efficiency compared to the Deep model (4 to 13 seconds vs 16 minutes to nearly 4 hours), making it the preferred choice for our case company.  Scoring individual log events for anomalousness seems like a good aid for root cause analysis of failing test cases, and our case company plans to add it to its online services.  
Despite the recent surge in using deep learning in software system anomaly detection, we found limited benefits in doing so. However, future work should consider whether our finding holds with different LSTM-model hyper-parameters, other datasets, and with other deep-learning approaches that promise better accuracy and computational efficiency than LSTM based models. 



\end{abstract}

\input{1_intro}
\input{2_background}
\input{4_results}

\input{5_related_work}

\input{6_conclusion}

\bibliographystyle{ieeetr}
\bibliography{ref}

\end{document}

%% file: 1_intro.tex
\section{Introduction}
Detecting anomalies in log events can be useful in long test sequences from stability testing. Often, problems can be non-catastrophic at first and take a
relatively long time to develop, ultimately
resulting in a crash after long run times. Testing for these types of problems
is complex, as the ultimate causes are not necessarily clear when the crash
happens.  Moreover, some of these problems often manifest as so-called
``Heisenbugs'', which are difficult to reproduce reliably. In contexts like
this, anomaly detection in log events can provide test and application
engineers with a means to simplify tackling these subtle bugs. Given a log for a
long-running application, identifying anomalous events within it can pinpoint
where things went wrong. Manually inspecting the logs is not an option since
they are massive, and most of the information contained therein is irrelevant to
the bug in question. However, anomalous entries in the log can give a
starting point for the root cause analysis of the observed issue.

Many papers in the literature use deep learning to detect whether a log sequence has anomalies or not e.g. 
\cite{logsy, logrobust, loganomaly, cnnlog, sialog}. 
Yet, only a few works \cite{deeplog, rosenberg2020spectrum} have focused on the root-cause analysis or failure localization in log sequences like we do in this paper. Recently there have been works on root-cause analysis or anomaly localization in Micro-service based systems, e.g. \cite{yu2021microrank, liu2021microhecl} where the goal is to find the failing micro-service among the services. 

Given the limited prior works and the company context, we follow an Agile software development principle "the simplest thing that could possibly work" \cite{beck2000extreme}. To this end, we chose two alternatives. A simple deep-learning approach using LSTM cells similar to what has been proposed in prior work \cite{deeplog}. To provide a comparison point, from classical computing we selected the n-gram approach. Both approaches have been used in analyzing sequences in natural language processing (NLP). 

\textbf{Insight:} In NLP context, deep learning has proven to be superior over n-gram but there are big differences in the text between natural languages and software logs. Natural language is semantically rich and the vocabulary is massive. Deep learning utilizes techniques like word embedding to compress the large and sparse word-document matrix to more compressed vector-document space and gains advantages over the n-gram approach. Yet, in software logs that lack the richness of natural language the advantage of deep learning might be mitigated. 


We define an anomalous log event as a log event that is unlikely to appear in the current context. Given a sequence of log events $E_1 E_2 E_4$ event $E_4$ is anomalous if $E_4$ rarely appears in the context where it is preceded by sequence $E_1 E_2$. Our work assumes that a sufficient amount of normal (not anomaly) execution logs are available, and they can be used to build anomalousness scores. We also assume that one knows which sequences are normal and which are not. These assumptions hold in the company case. 
We have four research questions. 
\begin{itemize}
    \item  RQ1: Which model is better N-Gram or LSTM model?
    \item RQ2: What is an optimal N-Gram configuration?
    \item  RQ3: Which model is better N-Gram or LSTM if we only consider accuracy over all configurations?
    \item RQ4: How can event anomalousness be measured and visualized?
\end{itemize}



%% file: 2_background.tex
\section{Company Case: Profilence}
This paper is collaborative research with a company called Profilence and the University of Oulu. The company wants to offer better investigation tools in its stability testing service. The university partners have an interest in understanding what approaches and algorithms are suitable for software log anomaly localization.

\begin{figure}[htb]
\centerline{\includegraphics[width=\linewidth]{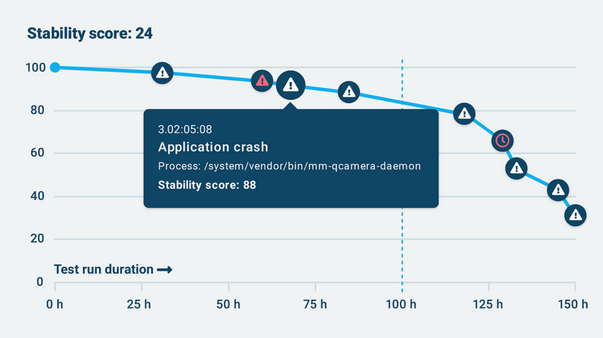}}
\caption{Stability test report}
\label{fig:stab}
\end{figure}

Profilence offers test automation and telemetry solutions with a focus on
long-term stability (reliability) testing on embedded devices. A common use case is to perform
long-running tests, in which a number of automated test cases are run on several
devices, over periods often lasting several days. Figure \ref{fig:stab} illustrates a stability testing report that Profilence's services provide to the their users. 
Among the data collected,
there are application and system logs. As can be expected, the amount of log
data collected is very large, and not really tractable by either the test
engineers or the application developers who want to identify and fix issues.

It is often the case that a particular test case may succeed a number of times,
and then fail. In those cases, identifying anomalies in the logs may provide a
good starting point for identifying the root cause of the failure, which may
precede the actual observed issue by a significant amount of time. Even in
cases where no crashes occur, other important non-functional metrics, such as
battery consumption or CPU usage can vary across software versions. In these
cases, detecting anomalous behavior (with respect to, say, a previous version's
baseline for the same test cases) can prove useful in identifying the cause for
the performance issues observed. A similar approach can be used for telemetry
data, but comparing logs across devices, as test cases do not exist as such in
that context.

In any case, given the amount of log data, these anomalies cannot be eyeballed,
or simply found by ``grepping'' through the logs, so an efficient and reliable
mechanism to detect them is highly desirable.

\begin{figure}[htb]
\centerline{\includegraphics[width=\linewidth]{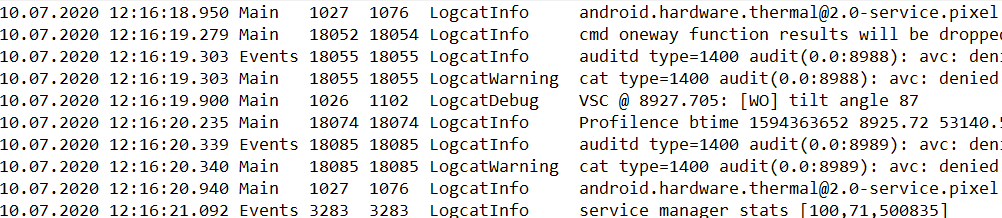}}
\caption{Snapshot of the Android testing log file}
\label{fig:snap}
\end{figure}

For a more concrete context, the Profilence data used for these experiments
correspond to a single test case for an Android camera application, see Figure \ref{fig:snap} for a snapshot. These tests
typically involve a number of automated steps performed on the application, from
opening it, maybe modifying some options, actually triggering a picture capture,
and so forth. A normal test run can span tens to hundreds of test cases such as
this, and can involve running them a very large number of times. In this
particular dataset, out of 275 executions of this test case, there are only two
runs where an exception occurred. The logs for this small data sample take about
768MB uncompressed. Typical test runs can easily generate one or two orders of
magnitude more log data, which indicates the need for automated analysis and
anomaly detection. The large amount of log data also motivates our investigation on the computational efficiency of the algorithms. 

%% file: 4_results.tex
\section{Results}
\subsection{RQ1 - N-Gram vs. LSTM}
\textbf{Background:}
Past work indicates that n-gram based models can offer relatively good performance in log sequence anomaly detection \cite{deeplog}. Regardless of this finding researchers' focus has recently been on using deep learning to log anomaly detection. Deep-Learning has been shown to offer state-of-the art performance but it is well known that deep learning models require high computational power. 

\textbf{Motivation:}
As little is known how n-gram models compare against deep learning, we first decided to investigate how it fairs against LSTM based deep learning model. After all, n-gram models are simple and transparent both of which are desirable properties when using the model in practice.

\textbf{Method:}
To investigate the difference an N-Gram model and a Deep Learning model were implemented. The Deep Model has an embedding layer with 50 dimensions to represent events, followed by two LSTM layers with 100 memory units followed by two deep layers with 100 units. This is architecture is very similar to \cite{deeplog} that has shown good performance in anomaly event localization. Based on the discussion with the company we decided to use four previous events to predict the fifth event (N-Gram with sliding window size n=5). 
Full implementation details and data for our work are available in the replication package\footnote{\url{https://github.com/M3SOulu/next_event_prediction}}. 


\textbf{Task:}
There are no datasets where anomaly events have been labeled. Log anomaly labels only exist in sequence level in existing datasets like HDFS \cite{hdfs}. Furthermore, for the company, it is too laborious to start labeling individual log events for each individual client. Thus, we need to use a surrogate task to understand how well different models would fare in identifying anomalous log events. To this end, we study the next event prediction task in normal sequences. We reason that if a model is good in predicting next events in normal sequences then the incorrect prediction that would occur in anomaly sequences can be considered as good candidates for being true anomaly events. We do not expect perfect next event prediction as it would mean that the system behavior is completely predictable by previous log events alone.

\textbf{Metrics:}
We measure accuracy, i.e. the share of correct event predictions among all predictions, and time taken for the computations. 

\textbf{Data:}
Table \ref{tab:data} shows descriptive statistics for our data. As with Profilence data, we are interested in long test sequences we select the normal sequences that have  10,000 events or more. There are 100 such sequences. Overall, these 100 sequences of Android camera testing have 4.5 million log events. To contrast the long log sequence of the case company (45,474  events on average), we use open data from HDFS \cite{hdfs} with 10,9 million log events from 588 thousand log sequences that have an average of 18.5 events per sequence. HDFS (Hadoop Distributed File System) is a module of the Apache Hadoop project. Another reason for using HDFS data is that it is very likely the most utilized software log anomaly testing date set making it the de facto benchmark data. Having two datasets with very different origins and properties gives our study some generalizability beyond the immediate interests of the case company. 

\begin{table}[htb]
\centering
  \caption{Normal data used for next event prediction}
  \label{tab:data}
\begin{tabular}{|l|l|l|}
\hline
                    & Profilence & HDFS       \\ \hline
Log Sequences       & 100        & 558,223    \\ 
Log Events (total)  & 4,547,451  & 10,887,379 \\ 
Log Events (unique) & 3,783      & 17         \\ \hline
\end{tabular}
\end{table}

\textbf{Pre-processing:} All logs are pre-processed by parsing the raw log lines, shown in Figure \ref{fig:snap}, to log events. We use state-of-the-art parser Drain \cite{drain} for log parsing. Log parsing assigns similar raw log lines to the same log event. So instead of operating with the raw text data we operate with log events like $E_1$, $E_2$, $E_3$ etc. After parsing and before model training, we add necessary padding events: start of sequence ($SoS$) and end of sequence ($EoS$). This is needed to be able predict the first event of a sequence and the termination of a sequence. For example with n-gram (n=3) the first n-gram could be $SoS\: SoS\: E_2$ and then the models need to predict $E_2$ given $SoS\:  SoS$. Similarly, last n-gram could be $E_5\: E_7\: EoS$ and then the models need to predict $EoS$ given $E_5\: E_7$. 
We split the normal data from both data sources to training and test set with a 50:50 split. 

\textbf{Computing Environment:} We run tests on two computing environments. First one is named Office PC that mimics typical off-the-shelf PC with AMD A8-7600 Radeon R7 3.1 Ghz, 4 cores, 16GB of memory and without GPU. Second one is named GPU PC with Intel Core Processor (Broadwell, IBRS) CPU 2,4Ghz, 28 cores, 224GB memory, and dual GPU NVIDIA Tesla P100 PCIe 16 GB. This was provided to us by CSC.fi IT Center for Science. This mimics a typical GPU virtual machine that are accessible through commercial cloud service platforms. Having GPU available is claimed to be very beneficial in reducing deep-learning training and inference time.  

\textbf{Results}
Table \ref{tab:ngram_lstm_pro} shows that the LSTM model has higher accuracy in the Profilence data. Only in 5 out of 50 test sequences is the N-Gram model better while in 45 the LSTM model is superior. If we compute overall accuracy in the test set, we see that the N-Gram model has an accuracy of 0.848 while the LSTM model is slightly better with an accuracy of 0.853. A Dummy model predicting always the most frequent event would get an accuracy of 0.476 so both models have clear predictive power. 

When we look at results from HDFS data in Table \ref{tab:ngram_lstm_hdfs}  we see that accuracy scores are almost identical between N-Gram and LSTM models. In 89\% of the sequences of the test set, the accuracy is the same for N-Gram and LSTM (it is a tie). 
Overall accuracy is almost the same with the N-Gram and LSTM models having accuracies of 0.849 and 0.846 respectively. A Dummy model always predicting the most frequent log event would get an accuracy of 0.154. 

Regarding computing efficiency in model training time, we can see that the N-Gram model is superior in both data sets and hardware environments. Training the N-Gram model takes between 3.9 and 12.8 seconds depending on the data set and the hardware. Training LSTM model takes from 10 minutes to nearly four hours. In an LSTM model training, there is a massive benefit of having a GPU available for training.  In the Office PC without GPU LSTM training takes nearly two and four hours in company and HDFS datasets respectively. With GPU PC the training times are only 11 and 10 minutes. In company data, the reduction in training time between Office PC and GPU PC is 90\% (10-fold improvement) and in HDFS it is over 95\% (20-fold improvement). Comparing Office and GPU PC in N-Gram model training shows that the GPU PC is faster with reduction of 31\% (1.4 fold improvement) in training time in both the company in HDFS data.

Inference time consists of predicting all log events in test data. Like in training also in inference time N-Gram model is superior. For the N-Gram model, prediction takes 2.6 and 8.6 seconds depending on the hardware and data. Inference with the LSTM model takes between 1 and 11 minutes. In LSTM inference we again find benefits of having GPU as the inference time is reduced by 75\% (4 fold improvement) and 87\% (nearly 8 fold improvement) when changing from Office PC to GPU PC.

\begin{table}[t]
\centering
  \caption{Results in Profilence data set. 
}
  \label{tab:ngram_lstm_pro}
\begin{tabular}{|l|l|l|}
\hline
                                                   & N-Gram & LSTM       \\\hline
Accuracy in test                                   & 0.848  & 0.853        \\
Wins in test data (tie = 0)                         & 5     & 45            \\\hline
Training time (all train data)    - Office PC                  & 5.7s   & 115min 12sec \\
Training time (all train data)    - GPU PC                & 3.9s   & 11min 20s \\\hline
Inference time (all test data)    - Office PC                  & 3.8s   & 7min 11sec  \\
Inference time (all test data)    - GPU PC                  & 2.6s   & 1min 50s \\\hline
\end{tabular}
\end{table}

\begin{table}[t]
\centering
  \caption{Results in HDFS data set.
}
  \label{tab:ngram_lstm_hdfs}
\begin{tabular}{|l|l|l|}
\hline
                                                         & N-Gram & LSTM       \\\hline
Accuracy in test                                         & 0.849  & 0.846      \\
Wins in test data (tie = 248,157) & 18,919 & 12,036     \\\hline
Training time (all train data) - Office PC               & 12.8s  & 230min 54s \\
Training time (all train data) - GPU PC               & 8.8s  & 10min 22s \\\hline
Inference time (all test data) - Office PC                        & 8.6s    & 8min 19s \\
Inference time (all train data) - GPU PC               & 7.1s  & 1min 4s \\\hline
\end{tabular}
\end{table}


\textbf{Discussion} 
Beyond the accuracy results, we need to consider the practical aspects of
implementing these approaches in the context of a product, which imposes certain
restrictions in the trade off between performance and resource consumption (e.g.,
compute time or storage). From the training and inference times observed
(cf. Table~\ref{tab:ngram_lstm_pro}), it seems clear that using the N-Gram model is the better
alternative of the two models when it comes to implementing anomaly detection
in production. Even when using a powerful GPU for LSTM, the N-Gram has an noticeable edge on computational efficiency, not to mention in infrastructure costs and energy efficiency. The high-end GPUs used in these tests retail for about 6000Eur (with VAT).  

As of this writing, the exact way in which this process will be
integrated into Profilence's products is not yet defined. A likely course of
action will be to trigger the analysis of log data on a case-by-case basis, for
example when an issue is detected in some test cases that have been successful
(or otherwise uneventful) in previous runs, or in different versions of the
software. This avoids the need to parse all logs, and keep the
datasets needed for the analysis at reasonable sizes, so that the training and
inference do not impose a burden on the system.

The N-Gram model is faster in training and inference than the Deep model because the N-Gram model is like a memory whereas the LSTM model is a computational model. Training the N-Gram model involves storing all n-grams of given lengths, their counts, and predictions. The LSTM-model, on the other hand, trains a neural network from the data which is slower than storing data to memory. The N-Gram model inference is just accessing the data that is stored using Python Dictionaries that have O(1) lookup time. LSTM-model inference on the other hand involves pushing a given n-gram through the trained neural network to get the prediction. Again such computations make the LSTM model slower. 

Although the N-Gram models performed faster than the LSTM model, the LSTM model could perform better at the inference time given a specialized production environment. Here we have calculated the duration of the inference by calling the Tensorflow's predict function. However, the model could be exported for the TensorflowServing environment. TensorflowServing enables high-performance computation, easy deployment, and out-of-the-box Tensorflow integration in production environments. 
Yet, the investigation of whether the LSTM model performance would improve if deployed to the TensorflowServing production environment is left as future work. 

\subsection{RQ2 - Best N-Gram Model Configuration}
\textbf{Background:} 
Given that the N-Gram model offered almost similar accuracy while being computationally far more efficient we decided to investigate what is the best N-Gram model configuration. 

\textbf{Method:} Unlike deep-learning or machine learning models that have numerous hyperparameters requiring sophisticated hyperparameter tuning, the N-Gram model in the basic version has only one parameter, which is the number of previous events to consider for the prediction of the next event. We do not consider n-gram smoothing or skip-gram models in this paper.  In the previous section, we evaluated the N-Gram (sliding window size n=5) model meaning we looked at four previous events to predict the fifth. Here we investigate what is the optimal value between n=2-10. The task, metrics, and data are as explained in the previous section.  

\begin{table*}[htb]
\centering
  \caption{Results in Profilence data set with N-Gram model}
  \label{tab:ngram_pro}
\begin{tabular}{|l|l|l|l|l|l|l|l|l|l|}
\hline
sliding window size n                                  & 2      & 3       & 4       & 5       & 6       & 7       & 8       & 9       & 10      \\ \hline
N-Gram Accuracy in test               & 0.659  & 0.715   & 0.763   & 0.848   & 0.843   & 0.837   & 0.831   & 0.823   & 0.815   \\ 
Wins in test data (tie = 0)    & 0      & 0       & 0       & 45      & 0       & 0       & 0       & 4       & 1       \\ 
Training time (all train data) - Office PC & 3.8    & 4.9     & 5.7     & 5.7     & 5.6     & 7.1     & 7.0     & 6.9     & 7.8     \\ 
Inference time (all test data) - Office PC & 2.6    & 3.0     & 3.6     & 3.8     & 3.4     & 4.1     & 4.0     & 3.7     & 4.3     \\ 
n-grams (unique)               & 48,511 & 112,773 & 182,960 & 253,302 & 320,457 & 384,203 & 445,695 & 504,794 & 561,889 \\ \hline
\end{tabular}
\end{table*}

\begin{table*}[!htb]
\centering
  \caption{Results in HDFS data set}
  \label{tab:ngram_hdfs}
\begin{tabular}{|l|l|l|l|l|l|l|l|l|l|}
\hline
 sliding window size n                          & 2     & 3     & 4     & 5     & 6     & 7     & 8      & 9      & 10     \\ \hline
N-Gram Accuracy in test           & 0.668 & 0.670 & 0.799 & 0.849 & 0.885 & 0.890 & 0.899  & 0.902  & 0.904  \\ 
Wins in test data (tie = 264,195) & 4     & 7     & 459   & 1,907 & 1,480 & 389   & 1,261  & 695    & 8,715  \\
Training time (all train data) - Office PC    & 8.7   & 10.9  & 11.6  & 12.8  & 14.1  & 15.2  & 15.9   & 15.9   & 18.5   \\ 
Inference time (all test data) - Office PC   & 6.9   & 8.3   & 9.5   & 8.6   & 10.3  & 9.0   & 9.1    & 11.9   & 9.9    \\ 
n-grams (unique)                  & 150   & 596   & 1,549 & 3,239 & 5,970 & 9,845 & 14.981 & 21,224 & 28,212 \\ \hline
\end{tabular}
\end{table*}


\textbf{Results:}
We can see that in Profilence dataset, see Table \ref{tab:ngram_pro}, the accuracy of the model increases until n=5 after which it start dropping again. However, in HDFS data the higher the n-value the better the accuracy. This can be considered surprising as HDFS sequences are short with average length of only 18.5 events. One could think that longer sequences in Profilence data would benefit from using the longer N-Gram models but this does not appear to be the case.

When we look at accuracy in individual test sequences we can see that in both datasets the most accurate models scores also most victories in individual test sequences. Yet, as other models also score victories this suggests that combining the N-Gram models with different lengths could offer even better accuracy. 

Looking at the training and inference time shows that increasing the N-Gram model length does increase both training and inference time. However, the increase is more modest than we expected. Going from n=2 to n=10 does roughly double the training time with 105\% and 112\% increase in training time in Profilence and HDFS datasets respectively. 

\textbf{Discussion:}
Before starting this investigation the company and the researchers were worried that the longer N-Gram models would hog up too much memory and offer poor computational efficiency. This is a reasonable intuition, as the company data has 3,783 unique log events, one would expect there to be a high number of bi-grams. The theoretical maximum is 14 million bi-grams ($3,783^2$) but in data only 48 thousand or 0.34\% exist. So what looked like a challenge in theory was not a problem in practice.

\subsection{RQ3 - N-Gram vs LSTM accuracy over all configurations}
\textbf{Motivation:} From the company viewpoint, the choice of going with N-Gram is clear from the results of RQ1 due to the much better computational efficiency and only slightly worse accuracy. Still, from the academic viewpoint, we wanted to look at how LSTM and N-Gram compare with each other with a different number of events used in prediction. 

\textbf{Method:} We used the same method as outlined in RQ1 and RQ2 but report only accuracy. 

\textbf{Results:} Table \ref{tab:ngram_vs_lstm_all_configs} shows that the N-Gram model has higher accuracy than the LSTM model when sliding window size n is between 2 and 4 in Profilence data. With longer sliding window when n is 5 or more the LSTM model is better. Actually, we notice that the N-Gram model has its peak performance when n=4 while LSTM improves the longer the sliding window size. 

For HDFS, the N-Gram model is always slightly better than the LSTM model.  In HDFS, both models keep improving as we use longer and longer sliding window size for prediction. 


\textbf{Discussion:} 
It appears that differences in accuracy between the LSTM model and N-Gram are small in our setup. The results suggest that LSTM would be a better choice when longer sequences from Profilence are analyzed. Also in Profilence data, we noticed that the LSTM model does better when sliding window size n used for prediction increases. In our study, we might have not used sequences that are long enough to highlight the benefits of the LSTM model. According to literature LSTM models can look back to 500 to 600 events \cite{hochreiter1997lstm}.  




\begin{table*}[!htb]
\centering
  \caption{Comparing N-Gram and LSTM accuracy over number (n) of events used}
  \label{tab:ngram_vs_lstm_all_configs}
\begin{tabular}{|l|l|l|l|l|l|l|l|l|l|}
\hline
n                          & 2     & 3     & 4     & 5     & 6     & 7     & 8      & 9      & 10     \\ \hline
Profilence N-Gram Accuracy in test  & 0.659  & 0.715   & 0.763   & 0.848   & 0.843   & 0.837   & 0.831   & 0.823   & 0.815  \\ 
Profilence LSTM Accuracy in test &  0.654 & 0.707 &	0.761 &	0.853 &	0.856 &	0.859 &	0.862 &	0.863 &	0.865\\ \hline
HDFS N-Gram Accuracy in test           & 0.668 & 0.670 & 0.799 & 0.849 & 0.885 & 0.890 & 0.899  & 0.902  & 0.904  \\ 
HDFS LSTM  Accuracy in test            & 0.663	& 0.667 &	0.797 &	0.846 &	0.880 &	0.886 &	0.894 &	0.897 &	0.900  \\ \hline
\end{tabular}
\end{table*}

\subsection{RQ4 - Metrics and Visualization for Anomaly Investigation}
\textbf{Background:}
Previous sections show that the N-Gram model can predict next events with good accuracy in normal data. Therefore, using it to flag potentially anomalous events in anomaly sequences seems like a promising approach

\textbf{Method:}
We designed two metrics to offer more information on the potentially anomalous events. We show a visualization of them and the company has evaluated them qualitatively. 

\textbf{Results:}
We designed two anomaly score metrics to measure event anomalousness. Just flagging an event to be an anomaly when the model fails to predict it correctly is not optimal. Using the correct prediction as normal and all other predictions as anomaly results in too many anomaly predictions. In past work \cite{deeplog}, this was solved by deciding that being among top-8 predictions makes event as normal and only beyond that, an event is as an anomaly. Although this improves the situation it still produces a binary value indicating whether an event is an anomaly or not. We think it is better to demonstrate event anomalousness as a continuous value. Our anomaly scores are an indication only that are used to reason about the root causes of the anomaly situation. 

The first metric we use is Occurrence Count which is the number of previous event occurrences in the given context. So for the N-Gram model (n=3) and an event sequence $E_1\: E_2\: E_4$, we calculate the past occurrences of the event $E_4$ in a given context which is the presence of the two previous events $E_1$ and $E_2$. This results in n-gram (n=3) raw counts. Second, we calculate Probability that a given event exists after the previous events. In our example case, it is the probability of the event $E_4$ given two previous events $E_1\: E_2$. To give a numeric example if event sequence $E_1\: E_2$ exists 100 times and event sequence $E_1\: E_2\: E_4$ exists 10 times, then occurrence count for the event $E_4$ is 10 and Probability is 0.1.   

\begin{figure*}[htbp]
\centerline{\includegraphics[width=\textwidth,height=\textheight,keepaspectratio]{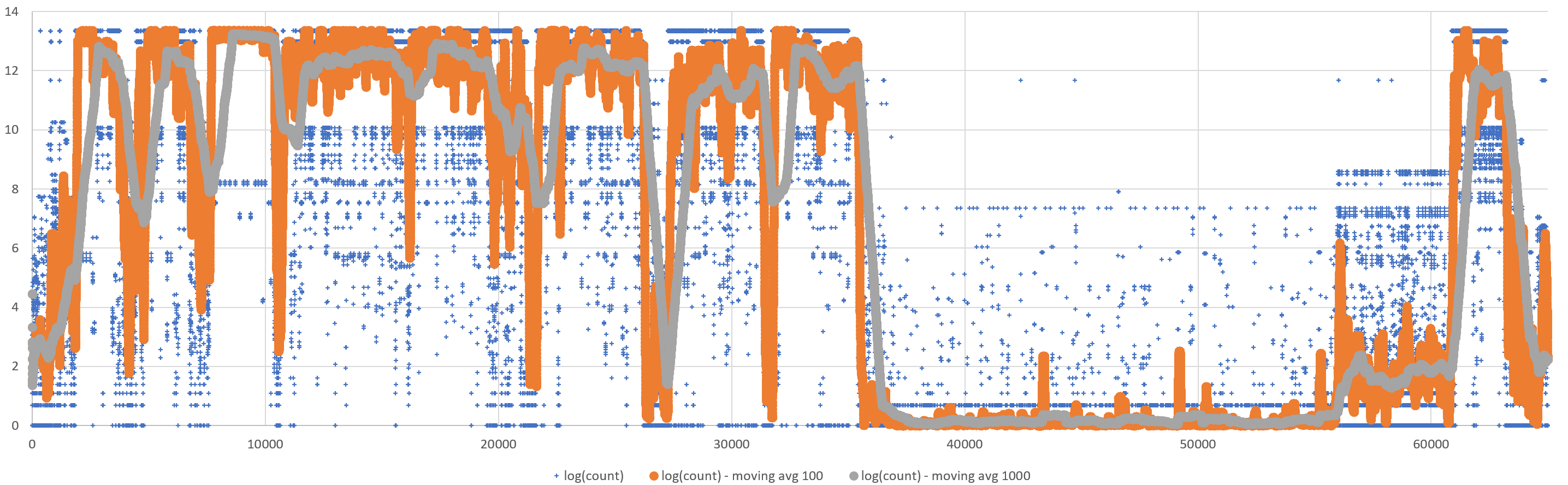}}
\caption{Occurrence count anomaly score log-scaled and moving averages over an anomaly sequence from Profilence with roughly 65,000 log events.}
\label{fig:count}

\centerline{\includegraphics[width=\textwidth,height=\textheight,keepaspectratio]{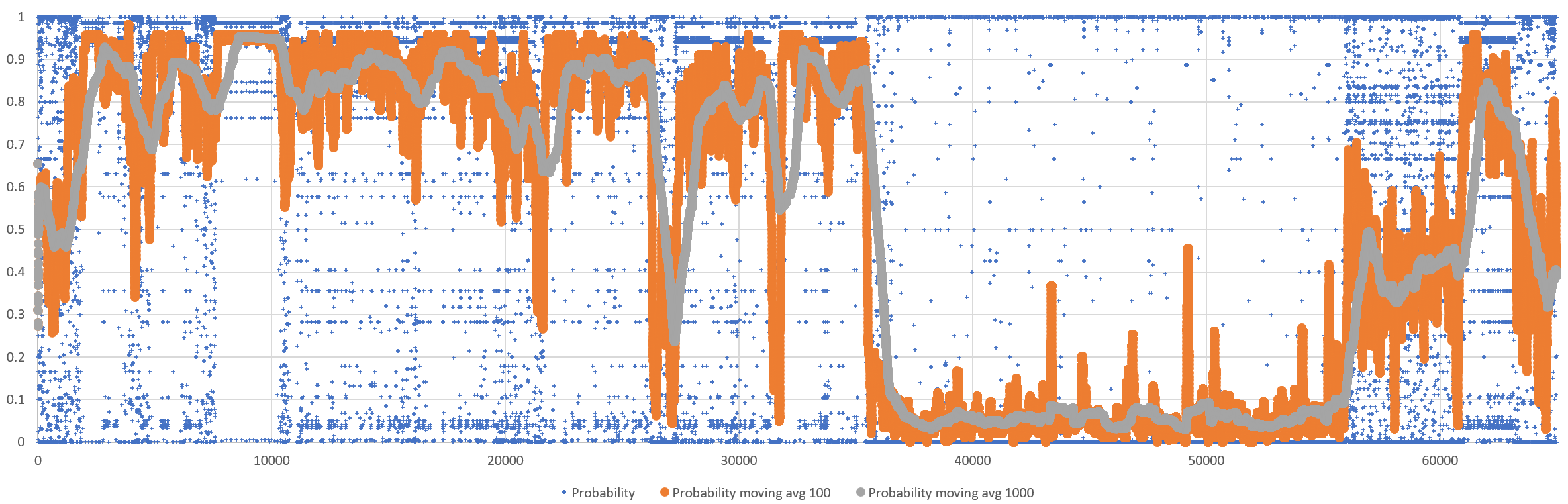}}
\caption{Probability anomaly score and moving averages over an anomaly sequence from Profilence with roughly 65,000 log events.}
\label{fig:prob}
\end{figure*}

Both metrics are shown in Figures \ref{fig:count} and \ref{fig:prob} with 
an anomaly case from Profilence with roughly 65,000 log events. The blue dots in the figures are scores for each event. We can see that individual event scores are quite noisy, more on that later. Therefore, we have added moving averages of 100 and 1,000 log event scores. From the moving average, one can more easily spot the different regions. For Figure \ref{fig:count} we have log scaled occurrence scores as otherwise the high values, e.g. 400,000 occurrences, would make it impossible to see differences in the lower values, e.g. 0, 10, or 100 past occurrences. Log scaling makes anomalous events that have the lowest scores more easily identifiable.

Having a view of where the anomalies in the logs are situated, the engineers
have a starting point for analysing the actual log data. While the anomalous
regions can span large parts of the logs, it seems sensible that the root causes
may lie close to the beginning of the anomalous data. As of this writing, it is
too early to tell how much of an improvement will the anomaly detection bring to
the bug identification and solving process, but the company expects that it will be an improvement over the status quo.

When comparing Figures \ref{fig:count} and \ref{fig:prob} one can notice that the graphs are highly similar. However, detailed event-by-event analysis shows differences between the scores.  Table \ref{tab:event_example} shows an example snippet of an anomaly event sequence from the HDFS data set. Raw log events have been omitted from the table to keep the presentation tidy but they are useful to the engineers in actual use.  On line number 14 we can see that both scores go down, 984 occurrences and probability less than 0.01. Then in line 15 past occurrences is still low at 977 but the probability of the event in line 15 is high at 0.99 given the previous events. We can see another drastic change at line number 22 where past occurrences go down only to 30 and probability goes down to 0.03. Finally, in line 28 we can see that past occurrences increases to 1,141, however, this is not reflected in the probability as it is already high in the previous events (lines 26 and 27). We use HDFS as an example as the changes in scores between individual log events are faster due to shorter sequence lengths. Yet,  similar patterns but with longer intervals between interesting changes also appear in Profilence data.

\begin{table}[htb]
\centering
  \caption{Example of event anomaly metrics (lower is more anomalous).}
  \label{tab:event_example}
\begin{tabular}{|l|l|l|l|l|}
\hline
Number & Event ID & Raw Event & Occurrences & Probability \\ \hline
12 & $E_{26}$ & [Omitted] & 429,147 & 0.93 \\ \hline
13 & $E_{26}$ & [Omitted]& 364,209 & 0.80 \\ \hline
14 & $E_{25}$ & [Omitted]& 984    & 0.00 \\ \hline
15 & $E_{18}$ & [Omitted]& 977    & 0.99 \\ \hline
16 & $E_{5}$  & [Omitted]& 1,216   & 0.99 \\ \hline
17 & $E_{16}$ & [Omitted]&676    & 0.49 \\ \hline
18 & $E_{6}$  & [Omitted]&1,047   & 0.98 \\ \hline
19 & $E_{26}$ & [Omitted]&1,071   & 1.00 \\ \hline
20 & $E_{26}$ & [Omitted]&1,141   & 0.98 \\ \hline
21 & $E_{21}$ & [Omitted]&1,074   & 0.94 \\ \hline
22 & $E_{18}$ & [Omitted]&30     & 0.03 \\ \hline
23 & $E_{25}$ & [Omitted]&46     & 1.00 \\ \hline
24 & $E_{5}$  & [Omitted]&48     & 1.00 \\ \hline
25 & $E_{16}$ & [Omitted]&25     & 0.52 \\ \hline
26 & $E_{6}$  & [Omitted]&93     & 0.98 \\ \hline
27 & $E_{26}$ & [Omitted]&93     & 1.00 \\ \hline
28 & $E_{26}$ & [Omitted]&1,141   & 0.98 \\ \hline
29 & $E_{21}$ & [Omitted]&1,074   & 0.94 \\ \hline
30 & $EoS$ & [Omitted]&1,074   & 0.94 \\ \hline
\end{tabular}
\end{table}

\textbf{Discussion:}
From Profilence's point of view, the current results seem promising, both in
terms of the approach working, and it being suitable for implementation in
production. At this stage, it is hard to quantify the impact it will have in
practice, but we expect it to simplify the root cause analysis of a broad class
of issues commonly found e.g., in Android systems. The next steps in this
process will involve running the N-Gram model on broader sets of data, and
defining a suitable implementation strategy within Profilence's analysis
pipeline, so that it can be tested by actual users, and its efficacy can be
better understood.

%% file: 5_related_work.tex
\section{Related Work}
A recent grey literature review in AI-based Test Automation identified areas of main test automation problems and AI solutions to these automation problems \cite{ricca2021ai}. Mapping our work back to that paper we can say that our work helps practitioners in the Test Closure area particularly in Costly results inspection and Visual analysis. In terms of AI solutions, we think our work fits the categories of Runtime monitoring.

A paper about test automation improvement in a DevOps team \cite{wang2020test} pointed out that test automation telemetry, i.e. the process of automatically collecting testing data to a centralized repository, was one of the success factors. We see that data collection through telemetry is a pre-requisite for our approach as we need data to form a model of the normal behavior. Such a model can only be built when a sufficient amount of data is available. 

From technical viewpoint, our approach is similar to Du et al. \cite{deeplog} that also operates at log event level rather than log sequence level. Our results are somewhat different and even conflicting. In HDFS open data set, they find that their N-Gram model has a lower F-Score than their LSTM model (0.94, vs. 0.96) whereas we find that our N-Gram model is slightly more accurate (0.904 vs 0.900). Considering the n-gram vs. LSTM results of both papers, we think that LSTM-based deep learning and n-gram have more or less similar performance and that the differences might be due to different tasks studied in the papers. However, where our results conflict is that Du et al. find that the N-Gram model offers the best performance looking back only to the previous event (sliding window size n=2). We find that the N-Gram model has its peak performance in HDFS when looking back nine previous events the N-Gram model (sliding window size n=10). Finally, \cite{deeplog} does not report any computing efficiency numbers that are important for our company case. 

Rosenberg and Moonen \cite{rosenberg2020spectrum} studied Spectrum-based method for aiding log diagnosis and root cause analysis using data from Cisco Norway. Their end goal is similar to ours but their method is quite different. Spectrum based log analysis applies techniques from source code fault localization research where the idea is to compute a suspiciousness score for each source code line based on its participation in passing or failing test case \cite{jones2002visualization}. In \cite{rosenberg2020spectrum} authors extend the same idea to log events. Their approach requires a large sample of log sequences of failing executions and thus does not apply to our company case. 

Pettinato et al. \cite{pettinato2019log} used Latent Dirichlet Allocation to study the behavior of a telecom system. Their intuition is that LDA topics consisting of different log events can be mapped to different system behavior, e.g. signal tuning or moving antennas. Their work tries to generalize behavior in a set of log events while we try to pinpoint focus to particular log events. Thus, the approaches are not alternatives but rather additive to each other. It would interesting to augment our log anomaly scores in Figures \ref{fig:count} and \ref{fig:prob} with labels of system behavior.

In anomaly detection of system calls, n-grams have frequently been used to detect anomaly sequences, e.g. n-gram counts become features in the prediction model that tries to predict if a sequence is an anomaly or not. These approaches have been used in different areas such intrusion detection \cite{khreich2017anomaly}, automotive \cite{rumez2020anomaly}, and aviation \cite{liu2018real}. However, those approaches operate on sequence level, i.e. is a particular sequence anomalous or not, while we operate on individual event level is particular event suspicious or not. 

Recently, Liu et al \cite{liu2020statistical} proposed a possibility to have a system-independent approach with frequency sequences of n-grams. From our viewpoint it would be interesting to see if it could also work on the event level as currently, all training in our approach needs to be done per system.





%% file: 6_conclusion.tex
\section{Conclusion and Future Work}

In this paper, we have addressed the topic of log anomaly root cause analysis or log anomaly failure localization. We compared a Deep-Model and an N-Gram model and found that the N-Gram model offers almost similar accuracy but much better computational efficiency in both long event sequences from Android camera testing and short event sequences from HDFS distributed file system.

Pinpointing anomaly events is seen as useful by the industrial partner that offers stability testing, aka reliability testing, services. The stability testing runs consist of tens of thousands of log events making it imperative to have automated support in suggesting suspicious events for investigation. Our approach with the metrics and visualization can be useful in debugging bugs that are difficult to reproduce or addressing flaky tests.

In addition, to the accuracy of anomaly predictions, the company values high computational efficiency. In this paper, we used data from 100 executions of a single test case. Typically a test suite for the company would contain tens or hundreds of such test cases from a product under test. Furthermore, the company offers testing for multiple products. Thus, the computational time needed to build a prediction model for all test cases in the company would be more than 1,000 times higher than what is presented in this paper, see Table \ref{tab:ngram_lstm_pro}. This heavily favors the model that has better computational efficiency. Sadly, many prior works investigating anomaly detection with deep learning do not provide measurements of computational efficiency leaving the reader guessing whether their approach would scale.  

Three important future work branches should be performed. First, we need to look at whether LSTM model performance can be improve with different model architecture and hyper-parameters. Also investigating the N-Gram model against other deep learning techniques like Convolutional Neural Networks (CNNs), BI-LSTMs, and Transformer is an interesting topic. Given the industrial desires to have models that are computational efficient the most promising choice would be to go with CNN models that are known to have good computational efficiency. 

Second, we need to see what improvements can be done to our N-Gram model. Skip-gram models seem like an obvious next step that might improve accuracy but unfortunately are likely to reduce computational efficiency. N-gram smoothing is often the user in NLP to address the problem that many n-grams are infrequent resulting in a sparse data matrix. We suspect this is less of an issue in software logs but still worth investigating.

Third, we need to study what is the impact of showing event anomaly scores to software engineers doing root-cause analysis. Currently, the company sees event anomaly scores as very promising but eventually it is the company's customers that determine its usefulness. 





